\documentclass[fleqn,usenatbib]{mnras}

\usepackage{newtxtext,newtxmath}

\usepackage[T1]{fontenc}

\DeclareRobustCommand{\VAN}[3]{#2}
\let\VANthebibliography\thebibliography
\def\thebibliography{\DeclareRobustCommand{\VAN}[3]{##3}\VANthebibliography}

\usepackage{graphicx}	
\usepackage{amsmath}	

\newcommand{\tess}{\textit{TESS}}
\newcommand{\swift}{\textit{Swift}}
\newcommand{\kepler}{\textit{Kepler}}
\newcommand{\gaia}{\textit{Gaia}}
\newcommand{\galex}{\textit{GALEX}}
\newcommand{\kms}{km\,s$\mathrm{^{-1}}$}

\newcommand{\Rsun}{$\mathrm{R_{\odot}}$}
\newcommand{\thalf}{$t_{1/2}$}
\newcommand{\energyone}{$3.2\pm1.4\times10^{30}$}
\newcommand{\energytwo}{$>1.7\pm0.7\times10^{31}$}
\newcommand{\energytwonosign}{$1.7\pm0.7\times10^{31}$}

\title[Optical and NUV flares from Ross 733]{\textcolor{black}{Characterisation of the NUV and Optical Emission and Temperature of Flares from Ross 733 with \swift\ and \tess}}

\author[J. A. G. Jackman]{
James A. G. Jackman$^{1}$\thanks{E-mail: jamesjackman@asu.edu (JAGJ)}
\\
$^{1}$School of Earth and Space Exploration, Arizona State University, Tempe, AZ 85287, USA\\
}

\date{Accepted 2022 September 30. Received 2022 September 30; in original form 2022 June 15}

\pubyear{2015}

\begin{document}
\label{firstpage}
\pagerange{\pageref{firstpage}--\pageref{lastpage}}
\maketitle

\begin{abstract}
We present the results of a coordinated campaign to simultaneously observe the M star binary Ross 733 simultaneously in the optical and near-ultraviolet (NUV) with \tess\ and \swift\ respectively. We observed two flares in the \swift\ NUV light curve. One of these was decay phase of a flare that was also detected with \tess\ and the other was only detected in the NUV. We used the \tess\ light curve to measure the white-light flare rate of Ross 733, and calculate that the system flares with an energy of $10^{33}$ erg once every 1.5 days. We used our simultaneous observations to measure a pseudo-continuum temperature of $7340^{+810}_{-900}$\,K during the flare decay. We also used our observations to test the NUV predictions of the 9000\,K blackbody flare model, and find that it underestimates number of flares we detect in our \swift\ NUV light curve. We discuss the reasons for this and attribute it to the unaccounted contributions from emission lines and continuum temperatures above 9000\,K. We discuss how additional observations are required to break the degeneracy between the two in future multi-wavelength flare campaigns.   

\end{abstract}




\section{Introduction} \label{sec:intro}
Increasing numbers of studies are now making use of photometry from wide-field exoplanet surveys to study the magnetic activity of low-mass stars. 
These surveys provide continuous and long baseline photometry of thousands of stars, allowing the estimation of starspot sizes, the measurement of flare rates and the study of flare energies as functions of stellar properties such as mass and age \citep[e.g.][]{Morris20,Yang19,Feinstein20}. However, by their design these observations cover only one part of the electromagnetic spectrum. Models must be calibrated using these observations in order to estimate the emission in other wavelength regions of interest \citep[e.g.][]{Rimmer18}. The predictions made are dependent on the chosen model. 
In order to get more in-depth information about the mechanisms at work during magnetic activity such as stellar flares, multi-wavelength observations are required. 

Coordinated multi-wavelength observations in the optical and the ultraviolet (UV) are essential for understanding the high-energy emission of low-mass stars. This is particularly relevant for stellar flares, rapid releases of energy due to magnetic reconnection events \citep[e.g.][]{Benz10}. These reconnection events accelerate charged particles from the upper stellar atmosphere down towards the chromosphere, where they impact the dense chromospheric plasma. When this happens, the nonthermal particles rapidly decelerate and release their energy into the surrounding area \citep[e.g.][]{Fisher85}. This energy release occurs at the footpoints of the magnetic field lines involved in the reconnection event. 
\textcolor{black}{The bulk of these particles are stopped in the chromosphere \citep[][]{Fletcher07}, and the energy deposited at these footpoints can heat the photosphere through methods such as backheating via chromospheric condensations \citep[][]{Allred06,Kowalski18} or Alfven waves \citep[][]{Russell13}, }
resulting in white-light emission \citep[e.g.][]{Hudson06,Watanabe17}. It is this emission that is detected with optical surveys such as NGTS, \kepler\ and \tess\ \citep[][]{Wheatley18,Borucki10,Ricker14}. Initial multi-colour measurements of the temperature associated with the continuum emission of white-light flares measured values around $\sim$10,000\,K \citep[][]{Hawley92}. However, more recent spectroscopic and multi-band photometric studies showed that flares exhibit \textcolor{black}{varying temperatures. \citet{Kowalski13} measured temperatures up to 14,000\,K at the peaks of flares from optical spectra. \citet{Howard20} used multi-colour optical photometry to measure values up to 40,000\,K, and \citet{Robinson05} reported a temperature of 50,000\,K for a flare detected in the UV with \galex.}   
These higher temperatures will result in stronger UV emission. However, while these optical studies can inform us about flare temperatures, they can not tell us about the relative contribution of the continuum to the total UV energy budget. UV observations of flares have shown that they have a significant contribution from emission features such as the Mg II, Si and Fe lines \citep[][]{Hawley07,Kowalski19}. Along with this, the UV is expected to have an elevated continuum due to the Balmer jump \citep[][]{Kowalski13,Kowalski19}. \citet{Hawley07} found that these emission lines can contribute up to 50\% of the flux. However, the exact fraction may depend on the energy and the morphology of the flare.  Simultaneous observations in the optical and UV can probe flare amplitudes and energies in both wavelength regimes, allowing us to test flare spectral models.

Space-based observations are required in order to study the UV emission of flares. 
One way of obtaining such observations is with the Ultraviolet/Optical Telescope (UVOT) onboard the 
Neil Gehrels \swift\ telescope \citep[][]{Roming05}. 
UVOT is equipped with a range of optical and near-ultraviolet (NUV) filters and two grisms, and has previously been used to obtain NUV photometry and spectroscopy of flares \citep[][]{Osten16,Wargelin17,Paudel21}. Simultaneous NUV and optical observations of flares can be obtained if \swift\ is scheduled to point at a target already being observed by a wide-field optical survey, such as \tess. \tess\ observes individual 24$\times$96 degrees sectors of sky for 27.4 days at a time, providing wide-field and long time-baseline optical photometry \citep[][]{Ricker14}. By pointing \swift\ at stars that are already planned for observation with \tess, we can obtain simultaneous measurements of optical and NUV flare light curves. These simultaneous UV and optical observations can be used to study the relative energies and amplitudes in each wavelength regime. 
\citet{Paudel21} recently used simultaneous \tess\ and \swift\ observations to study the emission of flares from the active M4 star EV Lac. They were able to measure the flare rate in both the \tess\ and \swift\ UVM2 bandpasses, and showed that \swift\ could be used to detect flares with energies lower than those typically detected by white-light studies. However, they did not detect any flares simultaneously with both \swift\ and \tess, stopping their analysis from studying how emission changed throughout an individual event. Therefore further observations with both \swift\ and \tess\ are required in order to study variations within flares. 

In this paper we present simultaneous observations of the Ross 733 system taken with \tess\ and \swift. Ross 733 is a spectroscopic binary comprised of two M stars \citep[][]{Shkolnik12,Koizumi21}. 
We have used the simultaneous optical and near-UV 
observations to study the flaring activity of Ross 733. 
We will present the methods used to reduce and analyse each dataset, and how we have combined them to study the activity of each star. In Sect.\,\ref{sec:results} we present the detection of two flares from Ross 733. One of these was detected with \swift\ only, providing an insight into which NUV flares may have white-light counterparts that are detectable with \tess. The other was observed with \tess\ and partially observed with \swift, allowing us to study the relative emission during the flare decay. 

\section{Observations}

\subsection{\tess}
Ross 733 was observed as part of \tess\ Sector 40 in Cycle 4, and has TIC ID 353011288. These observations were taken in the 2-minute cadence mode between the dates of 2021-Jun-24 to 2021-Jul-23 and include a total of 27.2 days worth of observations. In this work we have used the PDCSAP light curves, which have been detrended to remove instrumental effects and long-term drifts. Ross 733 was also observed in \tess\ Sector 14. However, to make sure our results are not affected by changes due to magnetic activity cycles in the two year gap between these sectors, we have chosen to focus on Sector 40 in our work. 

\subsection{\swift}
Ross 733 was observed with \swift\ between 2021-Jul-07 and 2021-Jul-24, for a total of \textcolor{black}{12.4 ks}. These observations were scheduled as a part of a Target of Opportunity request, to simultaneously study the optical and UV activity of Ross 733 during \tess\ Cycle 4 observations. 

Near-UV observations of Ross 733 were taken with the \swift\ UVOT instrument. We used UVOT to obtain NUV photometry using the UVM2 filter. This filter was used for all visits. The UVM2 filter covers wavelengths from 1699.08\,\AA\ to 2964.30\,\AA, with a central wavelength of 2259.84\,\AA. To analyse the \swift\ photometry we followed the method outlined in \citet{Paudel21}. UVOT observes in a TIME-TAG photon mode, with individual photon events being recorded. We first downloaded the raw data from the \swift\ archive, and then used \textsc{coordinator} and \textsc{uvotscreen} to convert coordinates to detector and sky coordinates, and then to create a cleaned event list. To generate a light curve for each visit we used the \textsc{uvotevtlc} tool with the cleaned event list. \textsc{uvotevtlc} requires the user to specify the source and background regions for aperture photometry. We used a circular aperture with a 10\arcsec\ radius for the source. We used a circular region with a 30\arcsec\ radius for the background, which was placed nearby the source. We used a time cadence of 11.033 s. Once we had generated each light curve, they were barycenter corrected using the \textsc{barycorr} tool. Finally, we converted each light curve to \tess\ JD.

\section{Methods}

\subsection{Ross 733}
In Sect.\,\ref{sec:intro} we noted that Ross 733 has previously been identified as a spectroscopic binary. This was first highlighted by \citet{Shkolnik12}, who measured the radial velocity (RV) change from $-88.8\pm1.1$ to $-73.0\pm0.8$ \kms\ in two observations. RV variations were later confirmed by \citet{Sperauskas19}, who measured RVs between $-76.5\pm1.8$ and $2.0\pm0.55$\kms, and \citet{Jonsson20} who measured RVs between -6.8 and -89.1\kms.

We plotted the position of Ross 733 on the Hertzsprung-Russell diagram using its astrometric information from \gaia\ DR2 \citep[][]{Lindegren18}. We found that it lies above the position expected for an isolated main sequence star. Instead, its position is consistent with that for an unresolved binary system, and is approximately 0.25 magnitudes above the position for an equal mass binary. This suggests that Ross 733 is comprised of two M stars of similar mass. To simplify our analysis we have assumed that Ross 733 is made up of two equal mass M3 stars. When we used this to account for the radius measured by \citet{Koizumi21}, we calculated a single star radius of $0.43\pm0.01$\Rsun. This value is inflated relative to model expectations for a single M3 star \citep[e.g.][]{Baraffe15}. Similar behaviour has been seen previously for M stars and has been attributed to convective inhibition and to starspots blocking photospheric flux \citep[e.g.][]{Feiden14,Jackson14}. We have used the effective temperature of $3367\pm39$,K measured by \citet{Koizumi21} from SED fitting, in this work. We have also adopted the distance of $18.09\pm0.02$\,pc measured by \citet{BailerJones18} from \gaia\ DR2 astrometry in this work.

\subsection{\tess\ Flare Detection and Analysis} \label{sec:tess_flare_detect}
To search for flares in the \tess\ light curves of each star we followed the method outlined in \citet{Jackman21kepler}, which was used to search for flares in \kepler\ 1-minute cadence light curves. This method employs a two step process, to first remove periodic modulation in the quiescent light curve due to starspots, and then to search for flares. To determine the period of any modulation we first used a generalised Lomb-Scargle periodogram. We searched each light curve for periods between 100 minutes and 10 days. If the most likely period had a power above 0.25, it was selected. We then smoothed the light curve using an iterative median filter with a window size of one tenth the selected period. We specified minimum and maximum window sizes of 30 minutes and 12 hours respectively. If the most likely period had a maximum power below the specified threshold of 0.25, a window size of 6 hours was selected. For a given light curve we performed our smoothing separately on each continuous segment. This was done to avoid jumps between different sectors which may alter the smoothed baseline. During each step of the smoothing process outliers lying more than 3$\sigma$ from the median were masked and interpolated over. This process was repeated iteratively until no more points were masked, or the process had run 20 times. We then divided the original light curve by the newly smoothed quiescent light curve. This resulted in a flare-only light curve, within which we searched for flares. To identify flare candidates in the detrended light curve we searched for consecutive outliers lying more than 6 MAD from the median of the detrended light curve. The MAD refers to the median absolute deviation, which is a robust measure of the variation in a dataset and has previously been used for flare detection \citep[e.g.][]{Jackman21}. We required flare candidates to have at least 3 consecutive outliers above this threshold.

We then visually inspected all flare candidates in our \tess\ light curves. We did this in order to remove potential false positive detections, such as due to asteroid crossings in the \tess\ postage stamps. We also manually set the start and end time of each flare during this stage. This was done to ensure we calculated the full energy of each event, as some flares had extended decays below the trigger level. 

To calculate the energy of each detected flare in the \tess\ light curve, we followed the method from \citet{Shibayama13} and assumed the flare spectrum was given by a 9000\,K blackbody. We equated the observed amplitude of the flare to the ratio between the flare and quiescent spectra in the \tess\ bandpass. We multiplied the energy of each flare by a factor of two to account for the dilution from each star in the binary.

We measured the flare rate following the method outlined in \citet{Jackman21}. The flare occurrence rate ($\nu$), the number of flares above a given energy $E$ per unit time, is expressed as:
\begin{equation}
\log{\nu} = C + \beta\log(E)    
\end{equation}
where $C$ is a normalisation constant and $\beta = 1-\alpha$. $\alpha$ is the slope of the power law used to describe the flare rate. 
This method uses flare injection and recovery tests to sample how the detection efficiency changes as a function of energy, and attempts to correct for this during the fitting of the flare rate. We injected \textcolor{black}{500} flares with amplitudes between 0.01 and 4, and \thalf\ durations between 2 and 60 minutes into the \tess\ light curve. These injected flares were generated using the \citet{Davenport14} model, which itself was created from 1-minute cadence \kepler\ observations of flares from GJ 1243. When fitting the flare rate we used an Markov-chain Monte Carlo (MCMC) process to fully sample the posterior distribution, using the \textsc{emcee} python package \citep[][]{ForemanMackey13}. We used 32 walkers for 10,000 steps and discarded the first 2000 as a burn in. 

\subsection{\swift\ UVOT Flare Detection and Energy Calculation} \label{sec:uvot_method}
To search for flares in the \swift\ UVOT light curves we used the method developed by \citet{Brasseur19}. \citet{Brasseur19} used this method to search for flares in \galex\ UV light curves \citep[][]{Million16}. This method searches for consecutive outliers in individual visits of a UV light curve, and calculates the start and end time as when the light curve goes above and returns to a calculated quiescence level. This method is designed to account for the short durations of the \galex\ visits. As individual \swift\ UVOT visit lengths are similar to those of \galex, we applied this method to the \swift\ light curves. Using an automated flare detection method also allows us to perform flare injection and recovery tests to test flare models, something we discuss in Sect.\,\ref{sec:testing_model}. We detected two flares using this method. 

We calculated the energy of each flare by multiplying the equivalent duration by the quiescent luminosity of the star in the NUV. The equivalent duration is a measure of the total flare emission, and is the time required for the quiescent star to emit the same amount of energy as a detected flare \citep[][]{Gershberg72, Hawley14}. 

To measure the quiescent luminosity of the star we took the average count rate of visits without any flares. We then converted this to the quiescent flux by multiplying it by the conversion factor of $8.446\pm0.053\times10^{-16}$ from the \swift\ UVOT calibration files. We then multiplied this by a factor of $4\pi d^{2}$, where d=$18.09\pm0.02$\,pc. This resulted in a calculated \swift\ UVM2 luminosity of 
\textcolor{black}{$2.73\pm0.04\times10^{28}\mathrm{\,erg\,s^{-1}}$} \textcolor{black}{for the total system, and $1.36\pm0.02\times10^{28}\mathrm{\,erg\,s^{-1}}$ per star, after accounting for there being two stars in the system.}. 
We also used our flare-free dataset to measure the average UVM2 magnitude of Ross 733. We measured an average UVM2 AB magnitude of \textcolor{black}{$18.17\pm0.02$ for the total system, and $18.85\pm0.02$ per star. }

\section{Results} \label{sec:results}

\subsection{\tess\ White-light Flare Rate}
\begin{figure}
    \centering
    \includegraphics[width=\columnwidth]{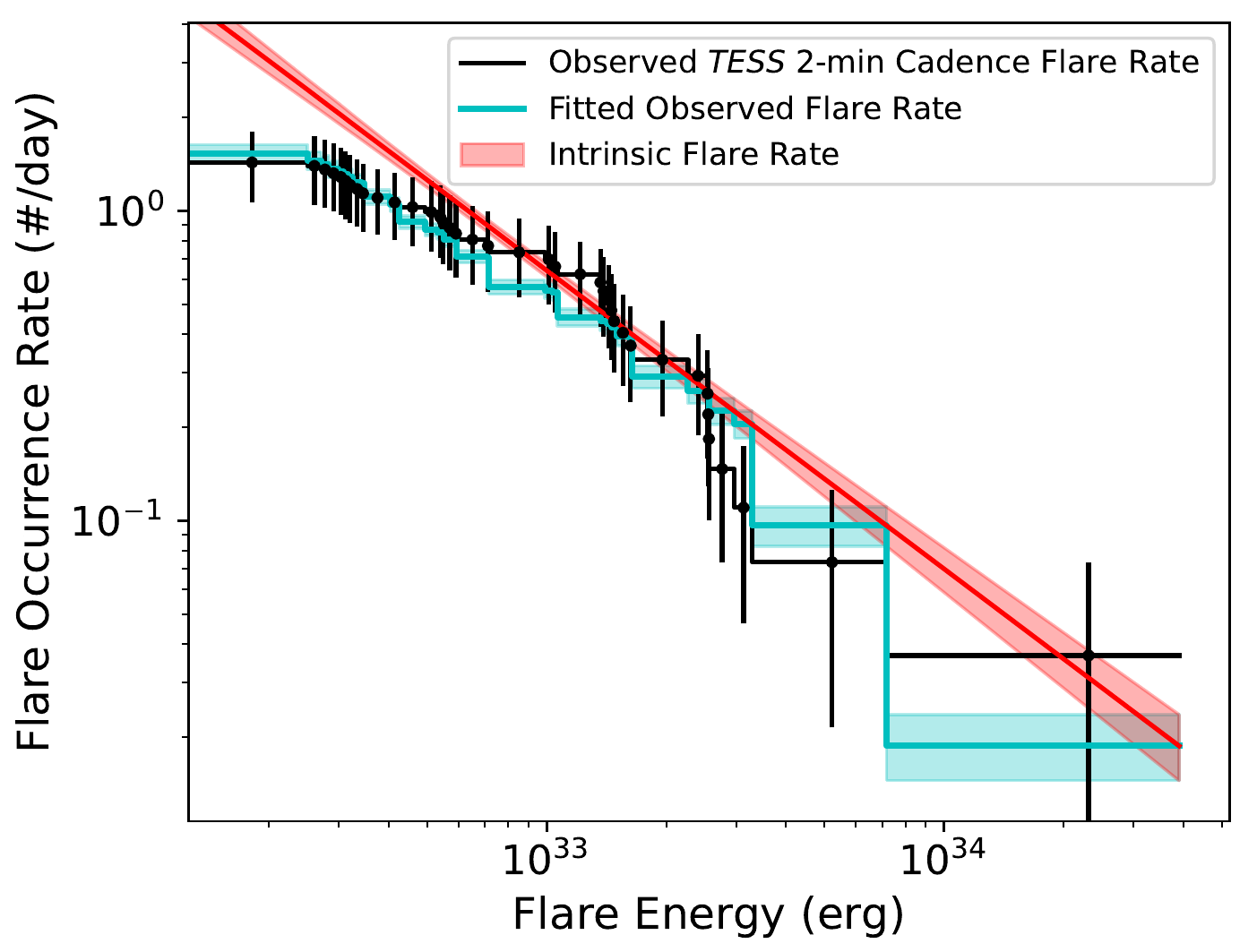}
    \caption{The white-light flare rate of Ross 733, as measured from \tess\ 2-minute cadence observations in Sector 40. We calculated the bolometric flare energies shown here using a 9000\,K flare blackbody model.}
    \label{fig:flare_rate}
\end{figure}

We detected 40 flares from Ross 733 in \tess\ Sector 40. These flares had energies between \textcolor{black}{$1.1\times10^{32}$ ergs} and \textcolor{black}{$3.9\times10^{34}$} ergs. 
We measured the white-light flare rate of Ross 733 following the method outlined in Sect.\,\ref{sec:tess_flare_detect}. The observed flare rate, and our best fitting model, are shown in Fig.\,\ref{fig:flare_rate}. 
We measured a best fitting flare rate from the full flare sample with $\alpha=$\textcolor{black}{$1.96\pm0.06$} and $C=$\textcolor{black}{$31.6\pm1.9$}. This corresponds to the Ross 733 system flaring with an energy of $10^{33}$erg and $10^{34}$ erg approximately once every 1.5 and 14 days respectively. This flare rate assumes that all the flares came from one star in the sample. If we assume the flares are split equally between the two stars, we calculate $C=$\textcolor{black}{$31.3\pm1.9$}. 
This corresponds to each star in Ross 733 flaring with an energy of $10^{33}$erg and $10^{34}$ erg approximately once every 3 and 28 days respectively.

\subsection{\swift\ UVOT Flare Analysis} \label{sec:uvot_results}
\begin{figure}
    \centering
    \includegraphics[width=\columnwidth]{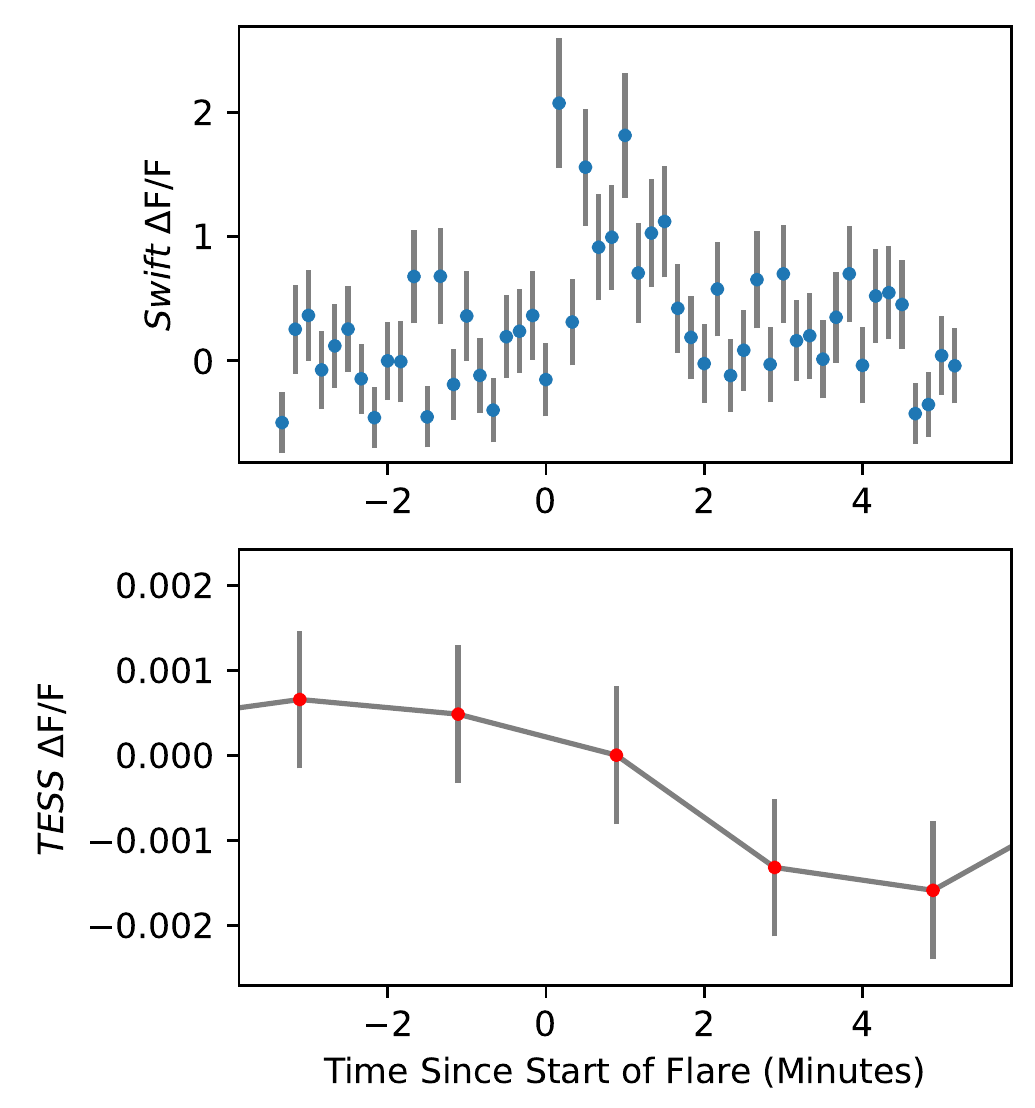}
    \caption{The first flare in our sample. The \swift\ light curve is shown in the top panel. The \tess\ 2-minute cadence light curve is shown in the bottom panel. This flare was not detected with \tess. We attribute this to its short duration, something we discuss in Sect.\,\ref{sec:uvot_results}. These are the original measured flare \textcolor{black}{light curves}, not adjusted for dilution.}
    \label{fig:first_flare}
\end{figure}

\begin{figure}
    \centering
    \includegraphics[width=\columnwidth]{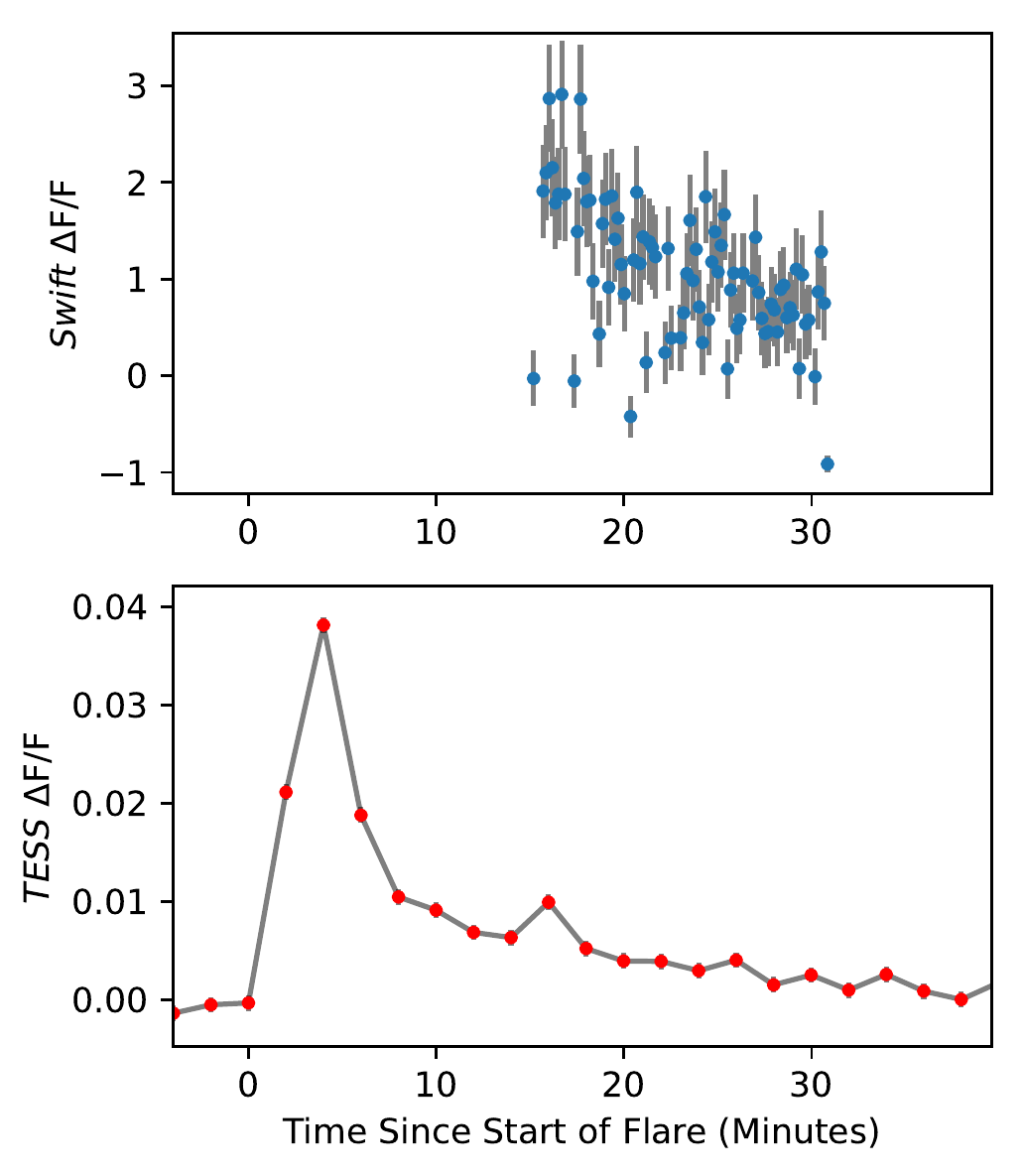}
    \caption{The flare simultaneously detected with \swift\ UVOT and \tess. The \swift\ observations began during the decay phase of the flare. Note the difference in the amplitude between the optical and the NUV. We calculated that this flare had an average pseudo-continuum temperature of $7340^{+810}_{-900}$\,K during the decay phase. These are the original measured flare \textcolor{black}{light curves}, not adjusted for dilution.}
    \label{fig:tess_swift_flare}
\end{figure}

We analysed the \swift\ UVOT light curve of Ross 733 following the method outlined in Sect.\,\ref{sec:uvot_method}. We identified two NUV flares. 
The first was a short duration flare which we only detected in the NUV. The second was a partial detection of a flare decay that we verified using the simultaneous \tess\ light curve. These flares can be seen in Fig.\,\ref{fig:first_flare} and \ref{fig:tess_swift_flare} respectively. The first flare had a total duration of 120 seconds. It lasted as long as one of our \tess\ exposures. This, combined with the reduced amplitude of the flare spectrum in the \tess\ bandpass may explain why we did not detect this event with \tess. The second flare lasted longer than the respective \swift\ visit, which lasted for 15 minutes. We measured a total duration of 70 minutes for this flare from visual inspection of the \tess\ light curve. 
We calculated the NUV energy of each event as \textcolor{black}{\energyone\ } and \textcolor{black}{\energytwo\ } ergs, with the lower limit corresponding to the partially observed flare. We can use this result to constrain the rate of NUV flares from the Ross 733 system with an energy above \textcolor{black}{\energyone\ }erg as \textcolor{black}{$14\pm10$ per day.} We have calculated the uncertainty on this occurrence rate assuming a Poisson rate for the flares. This estimated rate is less than the rate measured by \citet{Paudel21} for the active M4 star EV Lac. 

\subsection{Inferring Temperatures From A Simultaneous Optical and NUV Flare Detection} \label{sec:flare_teff}
During our analysis we identified one flare that had been partially observed in both of the \tess\ white-light and UVOT NUV light curves. Simultaneous detections of flares in the optical and NUV provide an opportunity to probe the underlying emission mechanisms. In particular we have been able to study the emission temperatures during the flare decay. We used our datasets to calculate pseudo-continuum temperatures. These pseudo-continuum temperatures correspond to the best fitting blackbody across the optical and NUV, and include the flux from blue optical and NUV emission features such as the Balmer jump and Mg II lines \citep[e.g.][]{Hawley07}. To calculate the best fitting pseudo-continuum temperature at each \tess\ and \swift\ observation we first generated a grid of blackbody curves with temperatures ranging between 5000 and 40,000\,K. We then binned our \swift\ light curve to match the \tess\ observing cadence. For each datapoint we normalised the pseudo-continuum blackbodies to all give the observed amplitude in the \tess\ bandpass. We did this using the PYPHOT python routine and the quiescent \tess\ magnitude of Ross 733, to avoid the use of a single star spectrum. We calculated the amplitude of each pseudo-continuum in the \swift\ UVM2 filter, using the UVM2 transmission curve\footnote{\url{http://svo2.cab.inta-csic.es/theory/fps/index.php?id=Swift/UVOT.UVM2&&mode=browse&gname=Swift&gname2=UVOT##filter}} and the quiescent UVM2 magnitude of $18.1\pm0.6$ calculated in Sect.\,\ref{sec:uvot_method}. We then interpolated our observed value on the grid of flare temperatures and synthetic calculated UVM2 amplitudes to calculate the corresponding pseudo-continuum flare temperature. We repeated this for each observation. When we used a pseudo-continuum to fit the total flux in the NUV, we calculated an average temperature of $7340^{+810}_{-900}$\,K during the flare decay. 
As mentioned above, spectra of flares from active M stars have shown that they exhibit strong NUV emission features and an elevated continuum due to the presence of the Balmer jump at \textcolor{black}{3646\AA} \citep[e.g.][]{Hawley91,Kowalski19} . \citet{Kowalski13} found that the Balmer jump alone could increase the continuum emission by between factors of 2 and 5. \citet{Hawley07} and \citet{Kowalski19} measured NUV line contributions of between 20 and 50 per cent the total flux. \citet{Kowalski19} also found that a 9000\,K blackbody curve, when normalised to the optical thermal continuum, underestimated the total NUV emission by a factor of 3. To try and account for the contribution of line emission and the elevated continuum, we calculated a second set of best fitting continuum temperatures. For this set we \textcolor{black}{multiplied the predicted NUV flux from the thermal blackbody by a factor of 3, to include the extra emission from the Balmer jump and emission lines.} 
When we did this, we measured an average flare decay temperature of $6150^{+560}_{-650}$\,K. 
These values are similar to those measured during the decay of flares by \citet{Kowalski13}. \textcolor{black}{We note that \citet{Kowalski19} fit the 9000\,K blackbody to the blue-optical continuum. The \tess\ bandpass covers redder wavelengths and may probe a component of the flare emission that has a different scaling factor than bluer wavelengths, such as the conundruum observed by \citet{Kowalski13}.}
Our measured temperature is also dependent on the assumed contribution from the UV emission lines. Our value of $7340^{+810}_{-900}$\,K can be considered a maximum value, however the fitted value is degenerate with the UV line flux. 
Future simultaneous optical and UV flare observations should aim to obtain either UV spectroscopy, to separate the line and continuum emission, or additional optical photometry in a separate filter to measure the continuum temperature \citep[e.g.][]{Howard20}. These steps would provide the extra information required to break the degeneracy and efficiently measure the contribution from UV line emission to flare energy budgets.  

\subsection{NUV and Optical Energy Partitions}
As part of our analysis in Sect.\,\ref{sec:results} we detected two flares in the NUV. One of these flares was not detected with \tess\ and had a NUV energy of \textcolor{black}{\energyone\ erg}. The other was partially observed and had a lower limit energy of \textcolor{black}{\energytwonosign\ erg}. We are able to use these results to probe the energy partition of flares in the white-light and NUV. For the partially observed flare, we recalculated the optical energy of the flare by isolating the times covered by both \tess\ and \swift. This resulted in a 9000\,K blackbody bolometric\footnote{\textcolor{black}{We note that this is the bolometric energy at the footpoint, and does not include X-ray or EUV emission associated with the corona \citep[e.g.][]{Osten15}.}} energy of $1.8\times10^{32}$ erg and a ratio of the NUV to bolometric energy of 0.09. We also performed this calculation assuming a flare temperature of 6150\,K during the decay. This temperature resulted in a bolometric energy of $1.2\times10^{32}$ erg and a NUV to bolometric energy ratio of 0.15. 

For our detection of a NUV flare without a white-light counterpart, we have used our flare injection and recovery tests from Sect.\,\ref{sec:tess_flare_detect} to determine which energies we are not sensitive to. We are unable, to a 3$\sigma$ confidence, to detect flares with bolometric energies corresponding to a 9000\,K blackbody below \textcolor{black}{$5\times10^{31}$ erg}. We calculated that this corresponds to a lower limit of the NUV to white-light energy ratio of \textcolor{black}{0.06}.

\subsection{Testing The NUV Prediction Of The 9000\,K Blackbody} \label{sec:testing_model}
We have used our simultaneous \swift\ and \tess\ light curves to test the UV predictions of the 9000\,K flare model used in our white-light energy calculations. \textcolor{black}{If the 9000\,K blackbody model can consistently describe the optical and UV emission, then we should be able to use our \tess\ observations to accurately predict the number and energies of flares in our \swift\ data.} 
The 9000\,K blackbody model has previously been used to predict the UV amplitude and energy of flares, by multiplying the bolometric flare rate by the fraction this model emits in a UV wavelength region of interest. We tested this model by calculating the UV flare rate corresponding to our fitted flare rate from Sect.\,\ref{sec:tess_flare_detect}. We followed the method from \citet{Brasseur19} to calculate that the 9000\,K emits \textcolor{black}{18.3\%} of its energy in the \swift\ UVM2 filter. We used this to calculate the predicted NUV flare rate of \textcolor{black}{the total} Ross 733 \textcolor{black}{system} as having \textcolor{black}{$C=30.9$ and $\alpha$=1.96}. 
We injected flares with energies corresponding to this power law directly into the \swift\ UVM2 light curve of Ross 733. We used the \citet{Davenport14} flare model to generate a grid of flares with full-width-half-maximum durations drawn uniformly between 30 seconds and 5 minutes, and amplitudes between 0.01 and 10 times the quiescent UVM2 flux. We performed injection and recovery tests 10,000 times. We used the results of these tests to calculate how often we detected two flares with an energy above \textcolor{black}{\energytwo\ erg}. \textcolor{black}{If over 68 per cent of our tests produced a light curve with two flares above our minimum measured energy, then the 9000\,K blackbody could feasibly provide both the optical and UV flare emission observed from Ross 733.} 
We found that 18 per cent of our tests \textcolor{black}{provided a light curve with a number and energy of flares consistent with our observed light curve.} 
This result implies that the 9000\,K blackbody does not provide enough flux in the NUV, however further tests with larger samples of NUV flares are required to determine how to correct for this. 
This result is in contrast to our measured flare temperature from Sect.\,\ref{sec:flare_teff}, which was below 9000\,K. However, this was measured during the flare decay. Previous studies have measured flare temperatures above 9000\,K during the peaks of flares, which are thought to drive the bulk of flare UV emission. Optical and FUV studies have frequently measured peak continuum temperatures up to 20,000\,K and in some cases 40,000\,K \citep[e.g.][]{Loyd18,Froning19,Howard20}. For a fixed amplitude in the \tess\ bandpass, a 40,000\,K blackbody will emit 25 times more \swift\ UVM2 flux than a 9000\,K blackbody. In addition to this, the contribution from emission lines as discussed in Sect.\,\ref{sec:flare_teff} will also increase the NUV flux above that predicted by a 9000\,K blackbody. However, NUV spectroscopy is required to resolve the relative flux contributions from the continuum and line features. 

\section{Conclusions}
We have presented the results of a multiwavelength campaign to study the magnetic activity of the Ross 733 system, using \tess\ and \swift. We have used the \tess\ light curve to measure the white-light flare rate of Ross 733, finding that the total system flares with a bolometric energy of \textcolor{black}{$10^{33}$ erg} approximately once every 1.5 days. We detected two NUV flares with \swift\ UVOT. These had NUV energies of \energyone\ and \energytwo\ ergs. The second flare was a partial detection during the decay phase. The second flare was simultaneously detected with \tess. We used the simultaneous detection to constrain the continuum temperature during the flare decay. We measured a pseudo-continuum temperature of $7340^{+810}_{-900}$\,K. When we incorporated the emission from UV lines, this temperature decreased to $6150^{+560}_{-650}$\,K. We also used our simultaneous light curves to test the NUV predictions of the 9000\,K flare blackbody model. We found from flare injection and recovery tests that this model underestimated the number of flares we expected to detect in the \swift\ data. We attributed this to peak flare continuum temperatures above 9000\,K and the contribution from line emission, neither of which are accounted for in this model. We also discussed how UV flare spectroscopy, or measurements in a second optical filter, are essential to resolve the degeneracy between the line and continuum emission. 
This work highlights how simultaneous optical and UV observations of flares can be used to constrain temperatures and test models. 

\section*{Acknowledgements}
We thank the referee for their helpful comments. We also thank Laura Vega and the \swift\ team for their help in reducing the \swift\ UVOT data. This research has made use of the SVO Filter Profile Service (http://svo2.cab.inta-csic.es/theory/fps/) supported from the Spanish MINECO through grant AYA2017-84089. JAGJ acknowledges support from grants HST-GO-15955.004-A and HST-AR-16617.001-A from the Space Telescope Science Institute, which is operated by the Association of Universities for Research in Astronomy, Inc., under NASA contract NAS 5-26555. This paper includes data collected by the TESS mission, which are publicly available from the Mikulski Archive for Space Telescopes (MAST). Funding for the TESS mission is provided by NASA's Science Mission directorate. This research was supported by the National Aeronautics and Space Administration (NASA) under grant number \textcolor{black}{80NSSC22K0125} from the TESS Cycle 4 Guest Investigator Program.

\section*{Data Availability}
This work is based on publicly available data which is available for download from MAST.



\bibliographystyle{mnras}
\bibliography{references} 




\bsp	
\label{lastpage}
\end{document}